\documentclass{emulateapj}
\usepackage{apjfonts}
\usepackage{amsmath}

\slugcomment{\sc To be published in ApJL}

\shorttitle{Evolved disk around a massive star?}
\shortauthors{Manoj et al.}

\begin{document}

\title{An evolved disk surrounding the massive main sequence star MWC~297?}

\author{P. Manoj \altaffilmark{1}}

\author{Paul  T. P. Ho \altaffilmark{1,2}}

\author{Nagayoshi Ohashi \altaffilmark{1}}

\author{Qizhou Zhang \altaffilmark{2}}

\author{Tatsuhiko. Hasegawa \altaffilmark{1}}

\author{Huei-Ru Chen\altaffilmark{3} }

\author{H. C. Bhatt \altaffilmark{4}}


\author{N. M. Ashok \altaffilmark{5}}

\altaffiltext{1}{Institute of Astronomy and Astrophysics, Academia Sinica, Taipei, Taiwan}
\altaffiltext{2}{Harvard-Smithsonian Center for Astrophysics, Cambridge, MA}
\altaffiltext{3}{Institute of Astronomy and Department of Physics, National Tsing Hua University, Hsinchu, Taiwan}
\altaffiltext{4}{Indian Institute of Astrophysics, II Block, Koramangala, Bangalore, 560 034,  India}
\altaffiltext{5}{Physical Research Laboratory, Ahmedabad - 380 009, India.}

\begin{abstract}
  
   We present the results of the interferometric observations of the
  circumstellar disk surrounding MWC~297 in the continuum at 230 GHz
  (1.3 mm) and in the (J=2-1) rotational transitions of
  $^{12}$CO,$^{13}$CO and C$^{18}$O using the Submillimeter Array. At
  a distance of 250 pc, MWC~297 is one of the closest, young massive
  stars (M$_{\star}$~$\sim$10 M$_{\odot}$) to us.  Compact continuum
  emission is detected towards MWC~297 from which we estimate a disk
  mass (gas+dust) of 0.07 M$_{\odot}$ and a disk radius of $\le$
  80~AU.  Our result demonstrates that circumstellar disks can survive
  around massive stars well into their main sequence phase even after
  they have become optically visible.  Complementing our observations
  with the data compiled from the literature, we find the submm dust
  opacity index $\beta$ to be between 0.1 and 0.3.  If the emission is
  optically thin, the low value of $\beta$ indicates the presence of
  relatively large grains in the disk, possibly because of grain
  growth. We do not detect any CO emission associated with the
  continuum source.  We argue that the $^{13}$CO emission from the
  disk is likely optically thin, in which case, we derive an upper
  limit to the gas mass which implies significant depletion of
  molecular gas in the disk.  The mass of this disk and the
  evolutionary trends observed are similar to those found for
  intermediate mass Herbig Ae stars and low mass T Tauri stars.

\end{abstract}


\section{Introduction}

There is now a growing body of evidence which indicates that the
formation of massive stars (8-20~M$_{\odot}$) is mediated by disk
accretion. Observational studies of massive protostellar objects in
the submillimeter(submm)/millimeter(mm) continuum and molecular lines
have revealed flattened disk like structures and outflows
perpendicular to them, suggesting that massive stars of early B or
perhaps even late O spectral types are born with disks
\citep[e.g.][]{zhang05, cesaroni07}.  If the high mass stars are
indeed born with disks, how long do these disks last around them? Do
they survive long enough so that planet formation processes can
proceed in them as is observed in the disks surrounding low and
intermediate mass stars?

Since the Kelvin-Helmholtz contraction timescale for stars more
massive than $\sim$ 8 M$_{\odot}$ is shorter than both the free-fall
and the accretion timescale (typically $\sim$ 10$^5$ yr), they arrive
on the zero-age main sequence (ZAMS) still embedded within their natal
cores and accreting from the surrounding circumstellar disk
\citep[e.g][]{mt03}.  Young high mass stars accrete a significant
amount of mass while on the ZAMS, even when surrounded by compact HII
regions \citep{keto02,sollins05}.  The intense UV radiation from the
central star eventually clears away the overlying envelope and the
star surrounded by a disk becomes visible in the optical/NIR
wavelengths.  Residual accretion from the surrounding disk on to the
central star may persist in these objects.  A few such stars (sp. type
B5 or earlier) have been known (e.g., Herbig Be stars) to show
classical spectrosocopic signatures of ongoing accretion and excess
continuum emission longward of 2 $\mu$m, indicating the presence of
circumstellar dust around them, presumably distributed in the
surrounding disks \citep[e.g][]{ngm00, manoj06}.  Direct imaging
studies at high angular resolution with interferometers have so far
detected compact continuum emission around two such objects, viz. MWC
1080, and R Mon, and evidence for Keplerian rotation in the optically
thick CO line emission from R Mon \citep{fuente03,fuente06}. However,
because of the large distances to these stars ($\ge$ 800 pc), even
with the high angular resolutions provided by interferometers, it is
not easy to distinguish between flattened structures of a few thousand
AU and {\it bona fide} circumstellar disks. 

In this letter we present the first interferometric observations of
MWC~297, which is a young 10 M$_{\odot}$ main sequence star of
spectral type B1.5 \citep{drew97}. Although distances of 450 to 870 pc
to MWC 297 have been cited in the literature \citep[cf.][]{canto84},
the more reliable distance estimate based on a detailed study of the
stellar properties and line-of-sight extinction by \citet{drew97},
places it at a distance of 250$\pm$50 pc.  MWC 297, thus, is one of
the closest, young massive star to us and is an ideal candidate for
high resolution studies.

\begin{figure*}
\plottwo{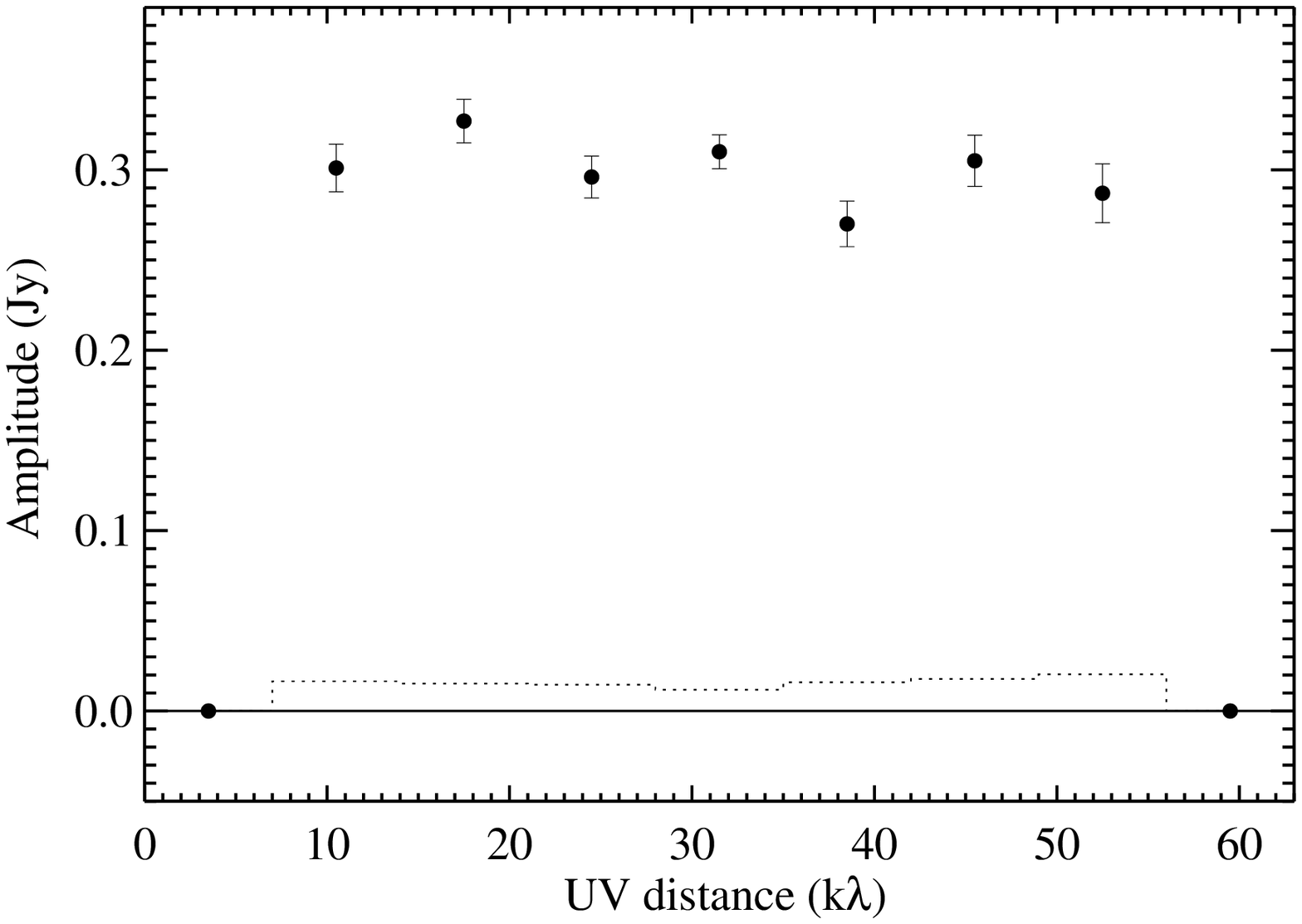}{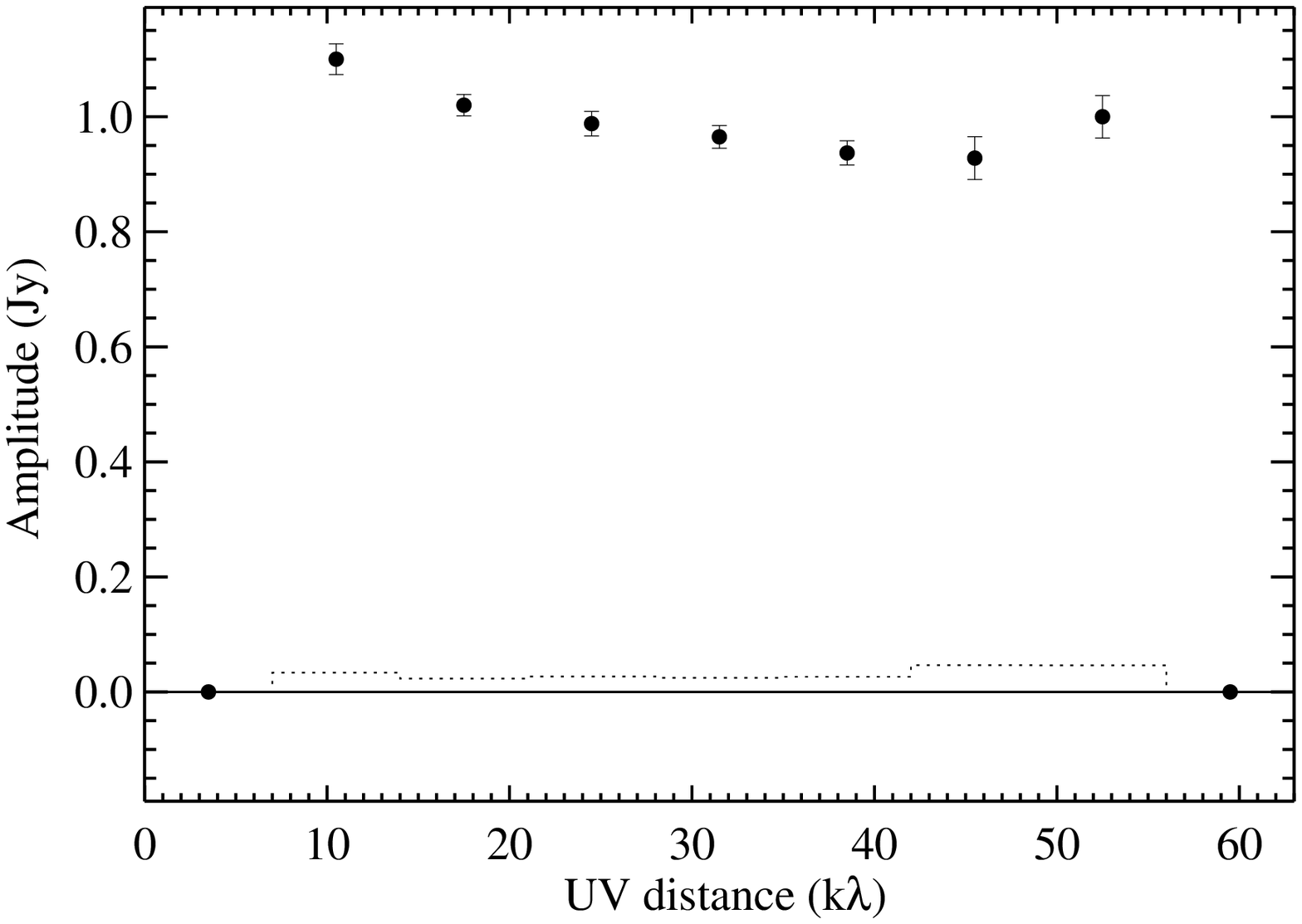}
\caption{Visibility amplitudes averaged vectorially in annular bins w.r.t the observed 1.3~mm continuum peak  plotted against the $uv$ distance for MWC~297 ({\it left}) and for the quasar 1743-038 ({\it right}). Error bars plotted are 1~$\sigma$ errors in amplitude.  The dotted lines indicate expected amplitude for zero signal.}
\label{vis}
\end{figure*}

\section{Observations \& data reduction}

MWC~297 was observed with the Submillimeter Array\footnote{ The
  Submillimeter Array is a joint project between the Smithsonian
  Astrophysical Observatory and the Academia Sinica Institute of
  Astronomy and Astrophysics, and is funded by the Smithsonian
  Institution and the Academia Sinica.}  (SMA) \citep{ho04} in the
continuum at 230~GHz (1.3 mm) on 28 August, 2006. The two sidebands,
each of 2~GHz bandwidth, include the rotational transitions (J=2-1) of
$^{12}$CO,$^{13}$CO and C$^{18}$O and the correlator was configured to
provide a velocity resolution of 0.5~kms$^{-1}$ for the CO~(2-1) line
which was centered in the upper sideband. The observations were made
in the compact array configuration of the SMA, where the projected
shortest and longest baselines were $\sim$~14~m and $\sim$~69~m (10
and 53~k$\lambda$) respectively.

\begin{figure}
\epsscale{1.0} 
\plotone{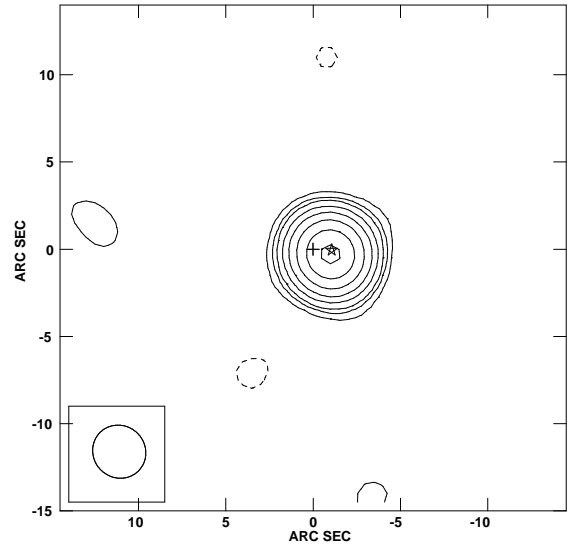}
\caption{Contour plot of 1.3~mm continuum emission observed towards MWC~297.  The cross corresponds to the phase center ($18^h27^m39.600^s$
  $-03^{\degr}49^{\arcmin}52.00^{\arcsec}$ J2000) and the star symbol
  denotes the stellar position from 2MASS. Boxed ellipse in the lower
  left corner indicates the position angle and FWHM of the synthesized
  beam. The contours begin at $\pm3\sigma$ (where $\sigma$ = 2 mJy
  beam$^{-1}$) and increase in steps of $\sqrt{3}$. }
\label{map}
\end{figure}

We used Uranus for both bandpass calibration and flux calibration.
Amplitude and phase calibrations were done with the quasar 1751+096
and the quasar 1743-038 was observed to verify the quality of phase
referencing from 1751+096.   The visibility data were calibrated using the MIR
package and the maps were generated and CLEANed using the NRAO AIPS
package.  Even after calibration the continuum visibility data of the
reference quasar 1743-038 showed evidence for phase decorrelation at
longer baselines.  The degree of decorrelation was similar for
1743-038 and MWC~297.  We self calibrated the visibility data for both
1743-038 and MWC~297 at a time interval of 4 min, which significantly
reduced the phase decorrelation.  The vector averaged visibility
amplitude plotted against the $uv$ distance for MWC~297 and the quasar
1743-038 after self calibration is shown in Fig. \ref{vis}.  We imaged
the self calibrated continuum visibilities of MWC~297.  The resultant
size of the synthesized beam was 3.$^{\prime \prime}$11 $ \times$
2.$^{\prime \prime}$97 with uniform weighting (PA $\sim$
45~$^{\degr}$). The contour map of the observed continuum emission at
1.3~mm is shown in Fig. \ref{map}. We expect the maximum uncertainty
in source positions to be $\leqslant$ $0.^{\arcsec}3$ based on the
positions of quasars mapped in our SMA observations.  Uncertainties in
the absolute flux is estimated to be 20\%.

\section{Results}

Compact continuum emission with a total measured flux density of
300~mJy is detected towards MWC~297 at 1.3 mm.  The angular separation
between the 1.3~mm continuum peak and the stellar position is less
than 0.$^{\arcsec}$3. Therefore, we assume that the compact continuum
source is centered upon the star. A two-dimensional Gaussian fit to
the continuum map of MWC~297 yields an observed size of 3.$^{\prime
  \prime}$14~$ \times$~3.$^{\prime \prime}$02 and a deconvolved size
of 0.$^{\prime \prime}$64$^{+0.^{\prime \prime}10}_{-0.^{\prime
    \prime}64}$~$\times$~0.$^{\prime \prime}$11$^{+0.^{\prime
    \prime}60}_{-0.^{\prime \prime}11}$ at
PA~=~165$^{\degr}$~$\pm$~15$^{\degr}$.  Allowing for the possibility
of the phase decorrelation in our observations affecting the observed
size, we take the longer dimension of the deconvolved size as an upper
limit to the source size. At the distance of MWC~297, it gives a
source radius of 80 AU.

MWC~297 is known to have an ionized wind associated with it
\citep{malbet07,drew97}.  The flux densities at 3.6 cm and 6 cm,
measured towards MWC~297 with the VLA \citep{skin93}, give a spectral
index of 0.6 (F$_{\nu}$ $\propto$ $\nu^{0.6}$), appropriate for
free-free emission from an optically thick, ionized wind. Assuming
that this emission continues to the mm wavelengths with the same
spectral index, we subtracted the possible contribution due to the
free-free emission from the total observed flux at 1.3~mm and obtained
a flux density of 200~mJy as due to dust emission.

If the dust emission is optically thin at mm wavelengths, assuming a
gas-to-dust mass ratio of 100 and a dust opacity per unit mass of dust
plus gas $\kappa_{\nu}$~=~0.005$\left[ \frac{\nu (GHz)}{230.6} \right
]^{\beta}$~$cm^2g^{-1}$ \citep{cesaroni07,kramer98}, we obtain a total
mass of 0.07 M$_{\odot}$ for the circumstellar material (gas+dust)
associated with MWC~297. A dust temperature of 100~K is assumed in
this calculation.  From the mass and the upper limit to the source
radius, we computed the optical extinctions along the line of sight
through the continuum source to the central star to probe the geometry
of the circumstellar material.  If the circumstellar material is
distributed in a spherical envelope of uniform density, the optical
V-band extinction would be $\ge$ 10$^4$~mag.  However, the observed
extinction $A_V$ is only 8 mag \citep{drew97}.  This strongly suggests
that the circumstellar dust is likely distributed in a comparatively
flattened and inclined morphology around MWC~297, perhaps in the form
of a disk.

\begin{figure}
\epsscale{1.25} 
\plotone{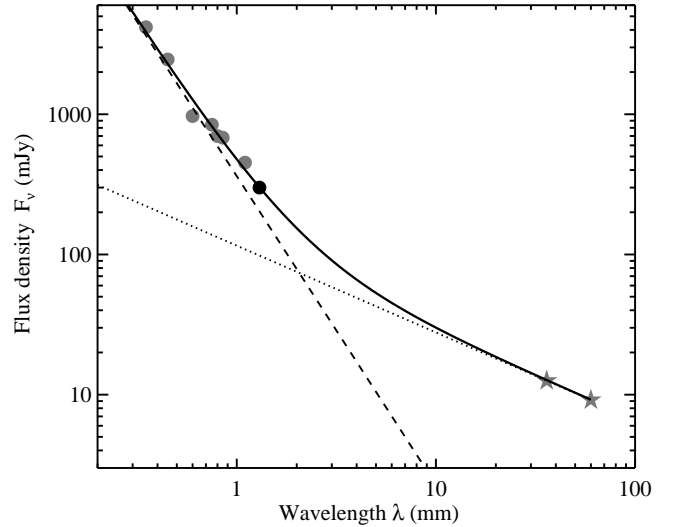}
\caption{SED of MWC~297 at cm, mm and submm wavelengths. The flux densities at cm  wavelengths (gray star symbols)  are from \citet{skin93} and at submm  wavelengths (gray circles)  from \citet{man94}. The black circle is our observed point at 1.3 mm. The dotted line is the model prediction for the free-free emission component from an ionized wind with $\alpha$ =0.6. The dashed line is the best-fitting disk model with disk mass of 0.07 M$_{\odot}$ and $\beta$ = 0.2. The solid line is the sum of both the components. }
\label{sed}
\end{figure}

In Fig. \ref{sed} we present the spectral energy distribution (SED) of
MWC~297 at submm, mm and cm wavelengths. We fit the observed points
with a combination of free-free emission arising in an ionized wind
and optically thin dust emission from a circumstellar disk of mass
0.07~M$_{\odot}$. The best fit is obtained when the dust opacity
power-law exponent $\beta$ has a value between 0.1 and 0.3.  This is
much lower than that observed for interstellar grains ($\beta$~=~2)
and the representative value used for circumstellar disks
($\beta$~=~1).  The low value of $\beta$ is often interpreted as
arising due to the average size of the emitting dust grains being
relatively larger.  Larger grain size in the circumstellar environment
around MWC~297 has also been reported independently from studies on
the wavelength dependence of extinction towards the star at the
optical wavelengths \citep{gortibhatt93}. This argues for possible
grain growth in the optically thin circumstellar material around
MWC~297.

In order to construct the SED and to determine the value of $\beta$ ,
we have used large beam ($\sim$ 20$^{\arcsec}$) single dish
measurements at submm wavelengths with JCMT in conjunction with our
interferometric data point at 1.3 mm.  The single dish measurements
are likely to have contribution from nearby extended structures.
However, major contribution to the single dish measurements still
comes from the compact continuum source that we detect with the SMA
and therefore, the derived value of $\beta$ should not be
significantly affected. For instance, at 1.3~mm, 60\% of the dust
emission observed with 19$^{\arcsec}$ beam of the JCMT \citep{man94}
and 74\% of the dust emission observed with 11$^{\arcsec}$ beam of the
IRAM \citep{henning98} comes from the compact continuum source of
$\sim$~0.$^{\arcsec}$6 in size.  Assuming that the relative
contribution from extended emission to the total dust emission
measured with JCMT is the same in all other submm wavelengths and
performing a similar fit yields $\beta$~$\lesssim$0.1.  Thus the value
of $\beta$ that we derive can safely be treated as an upper limit.

However, we note here that low value of $\beta$ is also consistent
with emission from an optically thick disk. If the 1.3 mm continuum
dust emission is optically thick, then the measured flux density
implies a disk radius of only $\sim$ 28 AU.  In order to have the
optical depth $\tau_{1.3mm}$~$\ge$ 1, such a disk should have a mass
M$_{disk}$ $\ge$ 0.06 M$_{\odot}$.

We do not detect any CO~(2-1) or $^{13}$CO~(2-1) emission with the SMA
at any V$_{LSR}$ at the position of the continuum source associated
with MWC~297.  The line of sight extinction of $A_V$~$\sim$ 8~mag
towards MWC~297 corresponds to CO~(2-1) and $^{13}$CO~(2-1) column
densities of about 9 $\times$~10$^{17}$~cm$^{-2}$ and
1.5~$\times$~10$^{16}$~cm$^{-2}$ respectively \citep{frerking82}, with
an assumed [$^{12}$CO]/ [$^{13}$CO] elemental isotope ratio of 60. The
optical depths of CO~(2-1) and $^{13}$CO~(2-1) will be 300/$\Delta
v$~(kms$^{-1}$) and 5/$\Delta v$~(kms$^{-1}$) respectively.  The
CO~(2-1) emission that we observe towards MWC~297 with the 10 m
Submillimeter Telescope (SMT) at ARO is extended, and has a velocity
width of $\sim$~10~kms$^{-1}$ (to be published elsewhere).  The
optically thick, extended foreground CO cloud is resolved out by the
SMA in our observations.  The $^{13}$CO~(2-1) line has a line width of
$\sim$~3~kms$^{-1}$ (SMT observations, to be published elsewhere). The
optical depth of the foreground $^{13}$CO will be 1 - 2.  If the
$^{13}$CO~(2-1) line from the circumstellar disk of MWC~297 has a
V$_{LSR}$ and a $\Delta v$ close to those of the foreground $^{13}$CO,
then most of the $^{13}$CO~(2-1) emission would have been absorbed by
the foreground component.  However, a Keplerian disk surrounding a
central mass of 10~M$_{\odot}$ will have a rotational velocity of
$\sim$~11~kms$^{-1}$ at a radius of 80~AU.  $^{13}$CO emission from
such a rotating disk would show a line width reaching 22 kms$^{-1}$
and we should have detected the high velocity component of the disk
emission away from the foreground cloud velocity.  Our channel maps
(velocity resolution = 0.5 kms$^{-1}$) at various V$_{LSR}$ in the
$^{13}$CO~(2-1) line indicate no hint of a localized $^{13}$CO
emission at the position of the dust continuum at an rms noise level
of 150 mJybeam$^{-1}$.

If the $^{13}$CO emission from the disk is optically thick, our
detection limit (53 mJybeam$^{-1}$/4 kms$^{-1}$) implies a source size
smaller than the dust disk size. This is contrary to what is observed
generally for cirsumstellar disks where the sizes of the gas disks are
a factor of 2-3 times larger than the dust disks
\citep[e.g.][]{simon00,pietu07}.  Moreover, the $^{12}$CO/ $^{13}$CO
line ratios observed for circumstellar disks in general suggest
$^{13}$CO emission to be optically thin \citep[e.g][]{thi01}.
Therefore the $^{13}$CO emission from the disk surrounding MWC 297 is
likely to be optically thin and from our sensitivity limit we obtain
an upper limit on the gas mass.  If the circumstellar material around
MWC~297 is distributed in a Keplerian disk of radius 80 AU and
assuming a gas excitation temperature of 50 K, [$^{12}$CO]/
[$^{13}$CO] ratio of 60 and H$_2$/$^{12}$CO conversion factor of
10$^4$, the upper limit on the gas mass estimated from our detection
limit for $^{13}$CO~(2-1) line emission is 3.5 $\times$ 10$^{-4}$
M$_{\odot}$.  This is about 200 times less than the disk mass
estimated from dust continuum emission and implies a gas-to-dust mass
ratio of $\sim$ 0.5.  Such discrepancies of about 2 orders of
magnitudes in the total masses derived from continuum and CO line
measurements have been known for disks around T Tauri stars and Herbig
Ae stars \citep[e.g.][]{mansar97, thi01}.  Suggested causes are gas
dispersal in the disk, CO depletion due to freezing-out onto the
grains and photodissociation, or the line emission being partially
optically thick \citep[e.g.][]{thi01,dut96}.  In the case of MWC~297,
which is a hot, main sequence star, gas dispersal and CO depletion due
to photodissociation are likely to be important.

\section{Discussion}
MWC~297 is a young, 10~M$_{\odot}$ main sequence star which is already
past the major accretion phase and has built up most of its mass. The
star is optically visible indicating that it has cleared most of the
surrounding material. Our interferometric observations with the SMA
show that it is still surrounded by a circumstellar disk of mass
0.07~M$_{\odot}$ and radius $\lesssim$ 80~AU.  The disk-like
structures detected in high-mass (M$_{\star}$ $\sim$ 8-20~M$_{\odot}$)
protostars have masses in the range of 1 - 12~M$_{\odot}$ and
radii~$\ge$~500~AU \citep[][and references therein]{cesaroni07,
  zhang05}.  The disk surrounding MWC~297, therefore, must be the
remnant of a more massive disk of the earlier phase, and this supports
the proposition that massive stars are formed via disk accretion. It
also demonstrates that circumstellar disks can survive around massive
stars well into their main sequence phase even after they have become
optically visible.  Reported evidence for the existence of Keplerian
disks surrounding 8~-~10~M$_{\odot}$ Herbig Be (HBe) stars also
support this assertion \citep{fuente06, schreyer06}.

Other than MWC~297, compact dust continuum emission from the
circumstellar disks surrounding optically visible early B-type stars
(HBe stars) has been detected only around two B0 type stars, viz.
MWC~1080 and R~Mon. Based on the disk masses derived from these
observations, it has been argued that the disk masses in HBe stars are
at least an order of magnitude lower than that in intermediate-mass
Herbig Ae stars and low-mass T Tauri stars \citep{fuente03,fuente06}.
However, because of the large distance to MWC~1080 (2200~pc
\citep{ngm00,manoj06}), the continuum emission detected at 2$\arcsec$
scale corresponds to a linear scale of $\sim$ 4000~AU and therefore
cannot be unambiguously considered as arising from a circumstellar
disk.  In the case of R~Mon, which is at a distance of 800~pc, the
continuum emission appears to come from a compact disk of radius
150~AU and mass 0.014 M$_{\odot}$ \citep{fuente06}. Thus the disk
masses estimated for early B-type stars are comparable to those found
for Herbig Ae stars \citep{ngm00} and T Tauri stars \citep{aw05},
given the uncertainties in the mass estimates.

The disk surrounding MWC~297 shows evidence for some degree of
physical evolution.  If the emission at submm and mm wavelengths is
optically thin, then the low value of $\beta$ that we derive implies
that the circumstellar dust grains around MWC~297 have sizes larger
than the interstellar grains. This indicates that the grain growth has
possibly begun in the disk.  Such grain growth has also been inferred
for the disk surrounding R Mon \citep{fuente06}.  We also find
evidence for significant depletion of CO in the disk surrounding MWC
297.  Interestingly, similar results indicating grain growth and
depletion of molecular gas in the disks have been reported for
low-mass T Tauri stars and intermediate mass Herbig Ae stars
\citep{aw05, acke04b, thi01}.  This would suggest that the
disk evolution proceeds not very differently in the disks surrounding
young massive stars from that observed for the disks in low-mass and
intermediate-mass stars.

Although the depletion of molecular gas and grain growth in the disk
seem to indicate that the disk surrounding MWC~297 is relatively
evolved, the spectroscopic signatures like a variety of emission lines
in the near-infrared and the optical wavelengths show that it is
driving a wind and is extremely active \citep[e.g][]{drew97,
  malbet07}.  \citet{malbet07} have successfully modeled the emission
lines as arising in a stellar wind with the high velocity
(600kms$^{-1}$) H$\alpha$ and H$\beta$ emission originating from a
large and somewhat spherical region while the Br$\gamma$ line is
confined to a narrow region just above the disk where the velocity is
dominated by the disk Keplerian rotation.  However, what drives this
wind is not very clear. Direct mass loss from the star and the
standard classical Be wind models have been found to be unsuccessful
in explaining the observed wind properties of MWC~297
\citep{malbet07,drew97}.  The activity in MWC~297 must be related to
the circumstellar disk surrounding it and it is likely that the wind
is accretion driven.  Though not very conclusive, spectroscopic
evidence seem to suggest that residual accretion persists in the
circumstellar disk surrounding MWC~297.

The disk that we detect around MWC~297 is possibly in an intermediate
evolutionary stage between the more massive accretion disk and the debris
disk.  The primordial disk material has begun to evolve and the disk
is in the process of being dissipated. It is interesting to note that
disks with similar observational characteristics in intermediate-mass
and low-mass stars are generally thought to be the sites of planet
formation. This opens up the intriguing possibility of planet
formation in the disks surrounding massive stars. However, detailed
studies of more such objects are needed to explore such a possibility.






\end{document}